\documentclass[cpp,fleqn]{w-art}
\usepackage{times}
\usepackage{w-thm}
\usepackage[]{graphicx}
\begin{document}
\{For the refereed version, see: DOI 10.10002/ctpp.201400073\}
\Volume{xx}
\Issue{x}
\Month{0x}
\Year{2014}
\pagespan{3}{}
\Receiveddate{21. September 2014}
\Reviseddate{ author's preprint}
\Accepteddate{1 Decmber 2014}
\Dateposted{ }
\keywords{Coulomb systems, plasmas, density functional theory, quantum many body theory.}

International Conf. on Strongly-Coupled Coulombo Systems (SCCS) 2014.


\title[Strongly-Coupled Coulomb Systems using DFT]{ A review of studies on Strongly-Coupled Coulomb Systems since the rise of DFT and SCCS-1977}


\author[Chandre D-W]{M. W. C. Dharma-wardana\footnote{Corresponding
     author: e-mail: {\sf chandre.dharma-wardana@nrc-cnrc.gc.ca}}\inst{1}} \address[\inst{1}]{National Research Council of Canada, Ottawa, Canada, K1A 0R6; 
and Dept. de Physique,  Universite de Montreal.}
\dedicatory{This Review is Dedicated to Leonard Ornstein and Frits Zernike on the Centenary of their equation.}
\begin{abstract}
The conferences on ``Strongly Coupled Coulomb Systems" (SCCS) arose from the  ``Strongly Coupled Plasmas" meetings, inaugurated in 1977. The progress in SCCS theory is reviewed in an `author-centered' frame to limit its scope. Our efforts, i.e., with Fran\c{c}ois Perrot, sought to apply  density functional theory (DFT) to SCCS calculations.  DFT  was then poised to become  the major computational scheme for condensed matter physics. The ion-sphere models of Salpeter and others  evolved into useful  average-atom models for finite-$T$  Coulomb systems, as in Lieberman's Inferno code. We replaced these by correlation-sphere models that exploit the description of matter  via density functionals linked to pair-distributions. These methods provided  practical computational means for studying strongly interacting electron-ion Coulomb systems like warm-dense matter (WDM). The staples of SCCS are wide-ranged, viz., equation of state, plasma spectroscopy, opacity (absorption, emission), scattering, level shifts, transport properties, e.g., electrical and heat conductivity, laser- and shock- created plasmas, their energy relaxation and transient properties etc. These calculations need pseudopotentials and exchange-correlation  functionals applicable to finite-$T$ Coulomb systems that may be used in  ab initio codes, molecular dynamics, etc. The search for simpler computational schemes has proceeded via proposals for orbital-free DFT, statistical potentials,  classical maps of quantum systems using classical schemes like HNC to include strong coupling effects (CHNC). Laughlin's classical plasma map for the fractional quantum Hall effect (FQHE) is a seminal example where we report new results for graphene. 
\end{abstract}
\maketitle                   





\section{Introduction}
The treatment of weakly interacting ions of charge $Z$ in electrolytes, in
classical plasmas and metals provided the impetus for Lorentz, Drude, Debye, H\"{u}kel, Onsager
 et al.,  to formulate the early theories of Coulomb systems.
 These were formulated as an expansion in a coupling parameter. One such
 parameter is $\Gamma$, viz., the ratio of the Coulomb
 energy to the kinetic energy. The Coulomb energy is of the order of $Z^2/r_{ws}$, where the Wigner-Seitz radius  $r_{ws}$ is the radius of the sphere containing one ion. 
The kinetic energy is, for classical
 systems,  simply the temperature $T$ (in energy units), while the Fermi energy is more appropriate for fermionic quantum systems. As long as
 $\Gamma <1$ the theory is satisfactory. However,  $\Gamma$ exceeded unity
for many systems of interest by orders of magnitude.
 Molten Aluminum contains Al$^{3+}$ ions and electrons,
 with $r_{ws}\sim 3$ atomic units. The corresponding ion-coupling parameter
 $\Gamma_{ii}$ is of the order of 1000 near the melting point! The coupling
 parameter $\Gamma_{ee}$ for electrons is of the order of 160.
 Even at very high temperatures, the system
 remains strongly-coupled for a large range of conditions, because the
 ionic charge $Z$ increases with $T$ from 3 to 13 when full ionization
 is achieved. Hence expressing physical properties as a series expansion
 in $\Gamma$, or within a hierarchy of equations based on $\Gamma$
 becomes non convergent. SCCS meetings recorded the progress in treating systems, classical or quantum, with coupling parameters larger than unity.


The electrons in electron-ion systems are quantum mechanical for a
 large range of conditions because of the high Fermi energy 
$E_F=0.5Z^{2/3}/(\alpha r_{ws}^2)$, where $\alpha=(4/9\pi)^{1/3}$.
 The system  is mostly in the `warm', strongly-correlated matter (WCM)
 regime since  $T/E_F \le 1$ and $\Gamma>1$. A sub-class of WCM is
 often known as `warm-dense matter' (WDM)~\cite{Graziani11}. The theory of WCMs
 involves finite-$T$ atomic physics,
 finite-$T$ continuum states inclusive of $i$-$i$,
 $i$-$e$ and $e$-$e$ correlations. 
 Several co-existing ionization types $\tau$ 
 with ionic charges $0\le Z_\tau \le Z_N$, and fractional compositions $x_\tau$
 can occur, where $Z_N$ is the nuclear charge.
  A mean charge $\bar{Z}=\sum_\tau x_\tau Z_\tau$ for an ``average ion" can usually be
defined and used in average atom-models to simplify calculations~\cite{MuriDW2013}.
 There are $s=\tau+2$  species of particles since electrons have
 two spin states. Treating spin types separately is needed to account
 correctly for exchange interactions and bound-states. Thus there
 are $s(s+1)/2$ coupling constants $\Gamma_{ss'}$ to deal with.
 Clearly, even when shell-structure is absent, perturbation methods, or
  truncations of hierarchies of equations (e.g., the BBGKY equations,
 Zubarev's equations of motion)
 cannot be used, unless some guidance from benchmark results is available.
 
The SCCS meetings since Dr. Gabor Kalman's inaugural meeting in 1977~\cite{Kalman77} 
have ranged from Europe, USA, Japan to Moscow~\cite{Moscow2005} and more, playing a pivotal role in bringing leading scientists together.
In the 1970s the SCCS-community studied simple `benchmark' systems
 like the one-component classical plasma (OCP)  described by
 the single parameter $\Gamma$.
 Benchmark molecular-dynamics (MD) results for the OCP were available
 for the free energy $F$,
 the equation of state (EOS), and even pair-distribution functions (PDFs).
 All thermodynamic properties and linear transport properties of systems
 at any $\Gamma$ can be expressed in terms of appropriate
 PDFs. The MD results served to establish the regime of validity of
  integral-equations like the Percus-Yevick (PY) and Hyper-Netted-Chain
(HNC) equations for the PDFs, when used with the celebrated 
Ornstein-Zernike equations~\cite{OZ14} whose centenary happens to be in 2014!
The Ornstein-Zernike equation, though simple in conception, forms a 
key-component of integral-equation  approaches used in the study of  strongly interacting systems.

In effect,  classical SCCS can be treated using classical
integral equations or molecular dynamics. However, theories of  
electron-ion systems need a tractable quantum treatment of the $e$-$i$ and $e$-$e$ interactions.

The traditional approach to the $N$-electron problem is to fix the nuclei
 in  given positions $R_i$ and solve the Schr\"{o}dinger (or Dirac) equation.
 This uses a many-electron wave function $\Psi$ expressed as a sum of all possible
 Slater determinants $D$ of rank $N$ made up of one-electron ``basis" functions,
 $N_b$ in number, with $N_b\gg N$. Each $D$ expresses an electron
 configuration for the fixed $R_i$. This approach to approximating
 $\Psi$ is known as configuration interaction (CI). The $e$-$e$
 interactions involve four (two incoming, and two outgoing) states.
 They give rise to $N^4$ Coulomb and exchange terms for each $D$.
 Furthermore (i) the CI problem grows in
 complexity as a non-polynomial function of $N$, 
even faster than the legendary hydra-head. (ii)
 The calculations have to be repeated for many sets of $R_i$ and
 a configuration average has to be taken.
 The Born-Oppenheimer approximation used in this approach is not valid
 when ion-electron coupled-modes are formed, or near the center of spectral
 lines where ion dynamics become important. (iii) For finite-$T$ problems,
 the number of angular momentum states $l$ needed in continuum-state basis
 functions of the type $\phi_{klm}(r)$, of positive energy $k^2/2m$ (in atomic units) rises very rapidly. Then even Hartree-Fock calculations
 become  prohibitive at finite-$T$.

This hopeless situation  was saved by the formulation of DFT in 1964
by Hohenberg and Kohn~\cite{DFT-source}. The historical background to DFT is given in an essay by Zangwill~\cite{ZangDFT14}. In DFT, the many-body $\Psi$  is dispensed 
with, and the ordinary density $n(r)$ in the
 fixed external potential (of the ions) determines
 the physics of the system. This requires knowing the kinetic energy as an
explicit  functional of $n(r)$. As this is not known, the
 practical implementation of DFT is via the Kohn-Sham method where
 the $n(r)$ at equilibrium is obtained from a Hartree-like
equation. The $N$-electron many-body
 problem is rigorously converted into an effective single-electron problem,
 and the many-body effects are buried in a one-body exchange-correlation
 potential $V_{xc}([n])$ whose existence is proved, but whose form is not
 available from DFT itself. The $T=0$ formalism was extended
 to finite-$T$ by Mermin where the free energy $F$ becomes a functional
 of the finite-$T$ one-body density $n(r,T)$. 

These developments were reported
 in the SCCS-conference series. Here we review
 them in the context of our work under the following topics:\\
(a) Generalization of the Hohenberg-Kohn-Mermin approach of $F[n]$ into a DFT theory encompassing {\it both} ions and electrons where $F([n],[\rho])$  depends on {\it both} the electron
one-body density $n(r)$ and the ion one-body density 
$\rho(r)=\sum_i\delta(r-R_i)$. This leads to {\it two} coupled variational equations~\cite{hyd0,SantaCruz}, one being a modified Kohn-Sham equation, while the other defines the variational equation for the ion distribution. 
(b) The construction  of exchange-correlation
 functionals $F^e_{xc}([n],T)$, and $F^e_{xc}([n],[\rho],T)$ for finite-$T$ Coulomb systems~\cite{excRPA,prl1,chncPRB,ppots12,Brown13}.
(c) The construction of two-temperature pseudopotentials
 and pair potentials for SCCS~\cite{drt,bep,ppots12}.
(d)  Their use for calculations of the equation
of state (EOS) over many orders of temperature and density~\cite{bep,pdw}.
(e) Calculation of transport properties, e.g., electrical and
 heat conductivity~\cite{res1,lvm}.
(f) Plasma spectroscopy, opacity (line widths, absorption, emission),
 and scattering~\cite{Grimaldi}.
(g) Laser- and shock- created plasmas, their energy 
relaxation and transient (optical and transport) 
properties~\cite{elr0,hazak,elr2001,elr-boundst}.
(h) Search for simpler computational schemes, for example,
classical potentials mimicking quantum effects, classical maps~\cite{chncPRL3d, chncPRL2d}, orbital-free methods etc. Their application to the  EOS, e.g., of hydrogen~\cite{chncHyd}, and 
 the fractional quantum Hall effect (FQHE)~\cite{fqheHNC1} in 
multi-component systems are presented, with some new results for graphene.

\section{Electron and ion one-body densities as the basis of the physical theory; an extended DFT.} 
\label{selfc}
In describing the physical properties of a many-body system, empirical
 models and  approximate interaction potentials
construed to give optimal
 results for  specific properties  usually fail for other properties.
 If the full many-body wavefunction $\Psi(\vec{r}_1,
\cdots,\vec{r}_N, \vec{R}_1,\cdots,\vec{R}_M)$ where $\vec{R}_i$ are nuclear positions is available, all properties can be consistently calculated from $\Psi$.
However, the $\Psi$ with or even without the dependence on $\vec{R_i}$ is not available, except for some small systems or model problems.

 Since the
 one-particle density $n(\vec{r})$ replaces $\Psi$ in DFT,  all
 properties become self-consistent if they are formulated as
 functionals of $n(\vec{r})$. All properties of systems interacting
via  a two-body potential (e.g., the Coulomb potential) can be formulated
transparently via pair-distribution functions (PDFs). A convenient way to
 relate PDFs to the one-body theory used in DFT is to
 to use a  system of coordinates with one of the particle (e.g., an ion)
 chosen as the origin  $r_0$. Then the particle correlations become manifest via the PDFs.
 The one-body density $n(\vec{r})$ becomes an effective two-body
 density $n(\vec{r}_0,\vec{r})$. For extended systems, such two-body densities
 relate simply to the corresponding pair-distribution functions
 $g(\vec{r}_0,\vec{r})$ and the mean densities $\bar{n},\bar{\rho}$ far away from the origin.
 Thus, assuming spherical symmetry typical of fluids, $\rho(r)=\bar{\rho}g_{ii}(r)$,
 $n_{ie}(r)=\bar{n}g_{ie}(r)$, and $n(r)=\bar{n}g_{ee}(r)$  hold. Hohenberg-Kohn theory implies that:
\begin{equation}
\frac{\delta F([n(r)],[\rho])}{\delta n(r)}=0; \;\;\;\;
\frac{\delta F([n],[\rho(r)])}{\delta \rho(r)}=0;
\end{equation}
In the usual formulations of DFT, only the first equation is considered
and the ions provide a ``fixed external potential".
The first equation is usually reduced to the Kohn-Sham 
equation, invoking an exchange-correlation potential acting on the electrons, 
while $\rho(r)$ is fixed. However, 
WCM calculations need a self-consistent evaluation of $\rho(r)$, to within the quasi-equilibrium time scales of the problem which may impose different temperatures to the different subsystems.
WCM calculations also need an extended form of xc-potentials 
which are functionals of both $n(r)$ and $\rho(r)$. That is, ideally, the applicable xc-potential has to be 
obtained {\it in situ}, as a non-local quantity  from a $g_{ee}(r)$ which is coupled to a calculation of 
$g_{ie}(r)$ and $g_{ii}(r)$. This is 
largely achieved in the Classical-Map HNC-technique (CHNC)~\cite{chncPRB} that will be discussed in  sec.~\ref{eos-H}.
 
 If DFT is valid,  the ONLY information required
 to describe (at least) the linear transport properties of the
 system  are just the equilibrium PDFs, be it a quantum system, or 
a classical or  mixed system, at $T=0$ or finite $T$, without requiring a wavefunction~\cite{cdwPisa2013,apvmm2013}. 
The exchange-correlation contributions are contained in the PDFs, and can be extracted using coupling-constant integrations~\cite{chncPRB}.
 The problem is to determine the PDFs
 for multi-component SCCS. However, once they are
 determined via DFT, the EOS, bound states, scattering phase shifts,  
static and dynamic conductivities, energy-relaxation rates, 
opacities, electromagnetic
 scattering etc., come out to be consistent with each other. 
Here we note that any PDF, i.e.,  $g_{ij}(r)$,  is connected with the
corresponding static structure factor $S_{ij}(k)$, while 
$S_{ij}(k\omega)$ determines spectra and time dependent properties. 
 
  This theory can be developed
 without making the Born-Oppenheimer approximation. The
 ion correlations in strongly-coupled systems  have to be mandatorily treated via a non-local ion-correlation theory since the local-density approximation fails badly for ions~\cite{hyd0}.

Hence, from the outset we  formulate SCCS problems in terms of PDFs and structure factors of the system,
 as reported in SCCS-meetings in Santa Cruz~\cite{SantaCruz}
 and subsequently. This approach
 contrasts that used in {\it ab initio} codes
 like {\it VASP} or {\it Ab-init}
 designed for condensed
-matter physics. In such codes the ion positions are fixed, and  specified as periodic repetitions of a simulation cell.
 The $T$=0 one-electron Kohn-Sham functions for a fixed ion 
configuration (Born-Oppenheimer approximation) are determined, and the 
 calculations are repeated for many static-ion configurations, to obtain averaged physical properties.  A thermal smearing of the one
-electron occupation numbers is introduced to take account of finite-$T$
 effects in a somewhat phenomenological way, and so far no serious attempts to use finite-$T$ versions of $V_{xc}[n]$ or finite-$T$ pseudopotentials have been made. 

The Kohn-Sham eigenfunctions are rigorously identifiable as states of the
 one-electron excitation spectrum only in the limit of negligible electron-
electron interactions. They are merely Lagrange
 multipliers in  the variational equations leading to the Kohn-Sham form.
 Nevertheless, they are (non-rigorously) used as valid single-electron
 states and excitation energies in, say, the Kubo-Greenwood formula to
 calculate transport properties of these systems. 
Alternatively, path-integral methods can be used, especially for hot systems
 where $T>E_F$. Quantum Monte Carlo is available at $T=0$,
 for small numbers of electrons. Owing to the finite size of the simulation
 cells, extrapolations to large-$N$ are needed, and results from different
 boundary conditions (e.g.,periodic versus twisted, etc. ) and different
 treatments of `backflow' can be dramatically different, especially with
 respect to exchange-correlation driven phase transitions which are difficult to predict~\cite{Drum2009,atta}.

\begin{figure}
\centering
\includegraphics*[width=12 cm, height=9.0 cm]{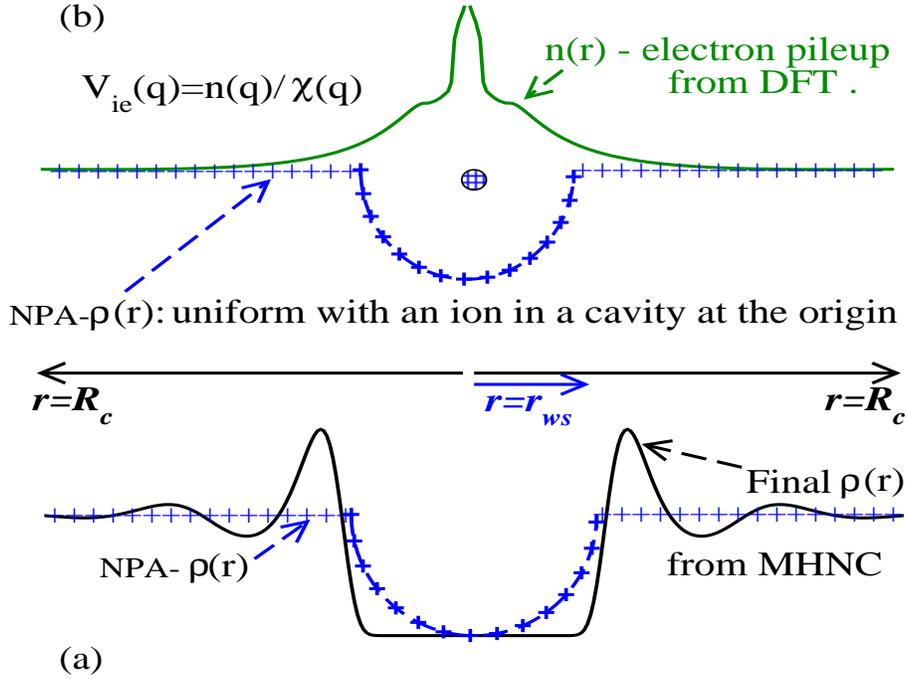}
\caption{ (a) A ``correlation sphere'' of radius $R_c\sim 6 r_{ws}$  defines the SCCS containing an ion distribution $\rho(r)$. This is
approximated in NPA with a spherical cavity for $r<r_{ws}$ and uniform $\rho$
for $r>r_{ws}$. (b)
 an electron distribution $n(r)$ around an ion at the origin is calculated 
from DFT for the given $\rho(r)$.
 The self-consistent
 $\rho(r)$ and $n(r)$ are determined by NPA for $n(r)$ followed by MHNC for $\rho(r)$.
Then pseudopotentials $V_{ie}(q)$, PDFs etc., needed for SCCS-physics follow from $n(r)$ and $\rho(r)$.}
\label{jelcav.fig}
\end{figure}

The approach based on PDFs lends naturally to
modeling the system within a `correlation sphere' (CS) built around a central nucleus,
encompassing all particle correlations. The CS, with a radius $R_c\simeq 5$-10 $r_{ws}$
encloses the ion distribution $\rho(r)$ and the associated
 electron distribution $n(r)$ forming a neutral system, labeled the `neutral pseudoatom' (NPA), introduced by Ziman as a name, and  formulated in detail by L. Dagens~\cite{Dagens73}. The idea is to approximately replace the interacting system of $N$ nuclei and their electrons with an equivalent {\it superposition of independent charge distributions} $n(\vec{r},\vec{R}_i)$ centered on each nucleus at $\vec{R}_i$, referred to as `neutral pseudoatoms'.  This is done in two
steps. The 1st step is to obtain $n(r)$, and the corresponding
 pseudopotentials $V_{ie}(r)$ and pair-potentials $V_{ii}(r)$.
The 2nd step  generates the fully self-consistent $\rho(r)$ and
 $n(r)$, as well as associated physical properties. The two steps are
depicted in Fig.~\ref{jelcav.fig}.
In the 1st step (the NPA), a fixed uniform ion distribution
 $\rho(r)=\bar{\rho}$ containing one nucleus in a spherical cavity of radius $r_{ws}$ is used to determine $n(r)$ 
 by a finite-$T$ Kohn-Sham calculation within the CS. This $n(r)$ is used to construct
$V_{ie}$, and $V_{ii}$.
(ii) Then a hyper-netted-chain (HNC) calculation using $V_{ii}(r)$ provides the full $\rho(r)$. Thus the ion correlations are given by the sum of hyper-netted chain graphs. Bridge-corrections to the HNC are needed for large-$\Gamma$ systems.The two-step implementation, involving first the NPA-DFT, and then the HNC for $\rho(r)$ can be merged into a single iterative scheme of a correlation-sphere DFT-HNC model, as in Ref.~\cite{hyd0}. But this is harder to converge.

A single nucleus is used as the `central object' in the NPA. Hence the method does not explicitly treat the formation of molecules and atomic clusters. The formation of H$_2$ molecules in a plasma would need calculations where H$_2$ is the central object. These are important at low densities and low temperatures. In such `cold-correlated low-density matter' situations, the central object has to be generalized to include a cluster of nuclei forming a ``neutral-pseudomolecule", rather than a ``neutral-pseudoatom".

We reviewed some of the early work of Singwi, Sch\"{o}lander, Tosi,
 Land, Ichimaru and others, and  the work of the Fermi-hyper-netted-chain
(FHNC) community in Ref.~\cite{prl1}. One of their aims
 was to calculate PDFs for SCCS,  but usually the resulting $g(r)$ had regimes 
with $g(r)<0$. The $g(r)$ from the CHNC method (sec.~\ref{eos-H}) is always positive, and provides a simple orbital-free classical approach to quantum calculations for extended condensed matter systems like Fermi liquids, plasmas, liquid metals and simple solids.
\section{Pseudopotentials and exchange-correlation functionals.}
\label{pseudopots}
A nucleus of charge $Z_n$ may carry $n_b$  tightly bound electrons in
 forming an atom or an ion, while the outer `valence electrons',  $Z=Z_n-n_b$ 
in number,  determine the main electronic properties of the
material. 
If many charge-states are present, one may use an average-ion
charge $\bar{Z}$ for an approximate treatment~\cite{MurDW2013}, but the full treatment is
straight-forward~\cite{pdw}. 
In simple models, the ions are regarded as point ions with charge $Z$ and
the core radius $r_c=0$,
  interacting with electrons by the
 Coulomb potential $V_{ie}(r)=-Z/r$, or $V_{ie}(q)=-ZV(q)$, with
$V_q=4\pi/q^2$.   Including the core with $r_c\ne 0$ gives the simplest pseudopotential. Although modern
pseudopotentials can be very sophisticated, 
they are set up for
standard condensed-matter applications, unsuitable for
general SCCS objectives. We find that 
simple generalizations beyond the point-ion model are sufficient
for plasma applications.  Thus the {\it local}
 Heine-Abarankov pseudopotential $V_{ei}(r)= A\theta(r_c-r)-\theta(r-r_c)Z/r$
where $r_c$ is finite and $A$ is the core potential, works very well. 
Here $\theta(x)$ is
the unit step-function, and the parameters $Z,A,r_c$ 
are  constructed from the free-electron component $n(r)$ of the
full Kohn-Sham density $n_{ks}(r)$, for the given 
 density and temperature range where $Z$ remains nearly fixed.
 However,
 the formulation of fully transferable {\it ab initio} finite-$T$ 
pseudopotentials even for a three-electron atom like Li is a very daunting task,
 taken up recently by Trickey
 and collaborators~\cite{Trickey}. 

\begin{figure}
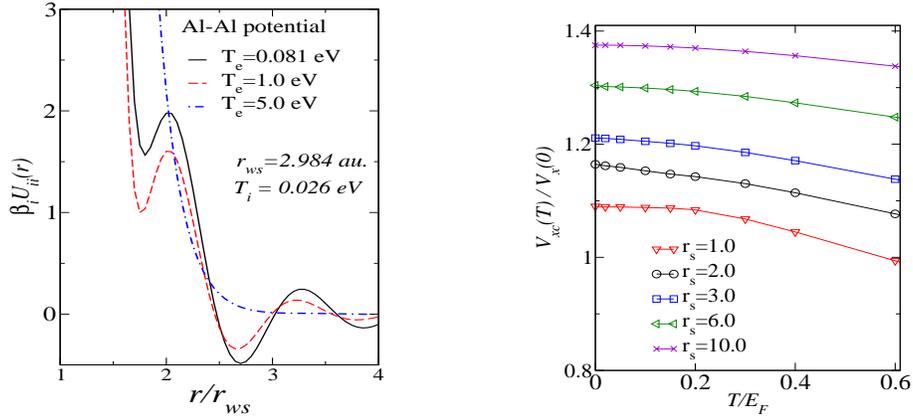

\centering
\vspace{0.3 cm}
\includegraphics[width=50mm, height=55 mm]{twoT-UiiAl.eps}
\hfil
\includegraphics[width=50 mm, height=52 mm]{xcpotsMS2.eps}
\caption
{(Left) Two-temperature ion-ion potentials for an Al-SCC-system
generated using NPA calculations.
(Right) Variation of $V_{xc}(r_s,T)/V_{xc}(r_s,T$=0) as a function of $r_s$
 and $T/E_F$) for a spin-unpolarized electron fluid (n.b. a previous version of this publication had
an error, with the y-axis marked $V_{xc}(T)/V_{xc}(0)$, instead of $V_{xc}(T)/V_{x}(0)$  and showed a spurious jump near $T=0$). }
\label{pairpot.fig}
\end{figure}

The free electron-pile up $n(q)$ in Fig.~\ref{jelcav.fig} is highly nonlinear. 
However, we attempt to make an effective  linear pseudopotential $V_{ei}(q)$ 
from the Fourier transform $n(q)$ and the linear response function
 $\chi(q;r_s,T_e)$ 
of the electron subsystem~\cite{ppots12}. Only the behaviour  in $0\le k<3k_F$ or some such range
(dependent on $T$) is used, since
the large-$k$ behaviour (the inner-core of the ion) is not needed.
When such a linearized approximation works, an
ion-ion pair-potential $V_{ii}(q)$ is also usually valid. Then (using atomic units),
\begin{eqnarray}
\label{pseud-pair.eq}
V_{ie}(q)&=&n(q)/\chi(q),\;\;\; V_{ii}(q)=Z^2V_q+|V_{ie}(q)|^2\chi(q)\\
\chi(q)&=&\chi^0(q)/\{1-V_q(1-G_q)\chi^0(q)\},\;\;\; V_q=4\pi/q^2
\end{eqnarray}
Here $\chi^0(q;r_s,T_e)$ is the finte-$T$ Lindhard function~\cite{KhannaGlyde76}, while $G_q$ is a
 $T_e$-dependent
`local-field' correction (LFC) available from the exchange-correlation energy
 or from $S_{ee}(k)$, as in Eq.~(13) of Ref.~\cite{prl1}, determined using CHNC. It is usually adequate to use the $q\to 0$ limit of $G_q$ when it is simply the compressibility ratio that can be accurately computed from the total energy (or equivalently from the exchange-correlation energy).
The $q\to 0$  limit is equivalent to the LDA in DFT.
An effective electron mass $m^*$ has been used in the electron response when appropriate~\cite{cdwAers83}. This $m^*$ differs slightly from unity due to band-structure and many-body effects. The parameters $Z, A, r_c$ in the pseudopotential
depend on the ion temperature $T_i$. Given $V_{ii}(r)$, the ion distribution
 functions
 can be calculated easily. Note that the Yukawa form of the pair-potential is
recovered for large ion-ion separations, at $T_e$ large enough to dampen
 the Friedel oscillations in the pair-potential (Fig.~\ref{pairpot.fig}).

Thus the NPA calculation provides the springboard to
constructing the whole system, be it solid or fluid, containing all the different ionization species
embedded together in equilibrium or quasi-equilibrium under the given
 conditions. This provides pseudopotentials, pair-potentials, correlation
functions, densities of states, and thermodynamic quantities like
 specific heats, compressibilities, phonon spectra  etc., without the need for
long calculations with large codes  generally used for
 solids. Seminal models of dense plasmas that used hard-sphere based calculations~\cite{MRoss79}, one-component plasma models, Thomas-Fermi theory, Yukawa potentials etc., can be replaced by NPA calculations that use physical potentials in a self-consistent manner.

Values for the modified electron chemical potential (`continuum
 lowering') as a function of density and temperature are relevant to both thermodynamics and spectroscopy.
 Experimental results for such `level-shift' data are becoming available only now~\cite{Vinko}, while the DFT-theory had been available for decades~\cite{pdw, kedge}.
Although DFT does not calculate energy levels, the difference in total internal energies between two configurations gives the energy of a spectral transition, and associated line shifts.

{\bf Controversial views regarding $\bar{Z}$.} \\
Some authors have  wondered if the concept of mean ionization 
has any physical meaning in SCCS.  The simple Saha equation is not applicable, but a valid thermodynamic or response-function formulation for $\bar{Z}$ in SCCS exists when there are delocalized electrons~\cite{pdw}. The $q \to 0$ limit of the electron-electron density response identifies  a plasma frequency which defines an effective free-electron density. This is in fact the electron density $\bar{n}=n(R_c)$ at the edge of the `correlation sphere' (i.e., $r=R_c$ in Fig.~\ref{jelcav.fig}) where the ion density is $\bar{\rho}=\rho(R_c)$. Hence, $\bar{Z}=n(R_c)/\rho(R_c)$. For self-consistency, this $\bar{Z}$ should agree with the ion-charge appearing in the pseudopotential, while $n(R_c)$ should agree with the $n_e$ appearing in the Drude conductivity and the electron specific heat. As discussed  in Ref.~\cite{dwp-hopping92} and elsewhere~\cite{pdw}, there is a difficulty in density regimes where there may be a plasma-phase transition; or at the threshold of a bound state changing into an ionized state. In such cases, a `hopping-electron' description (as in the 'mobility edge' of disordered semiconductors) may be appropriate. In such situations, the average atom-model has to be replaced by explicit consideration of co-existing ionization states and plasma phases, and possibly including localized electron states on ion clusters.  

 $\bar{Z}$ can be determined experimentally for a plasma of matter density $\rho$, from its conductivity $\sigma(\omega)$  near $\omega\to 0$, i.e., far from any interband absorption, where the Drude formula is valid.  Both the free electron density $n_e$ and the effective mass $m^*$ of the electrons can be obtained from the Drude absorption, as done for solid-state plasmas in metal films and laser-generated plasma slabs~\cite{Theye70, Ping-Ng06}. Then $\bar{Z}=n_e/\rho$. Most solid state plasmas have an $n_e$ such that $2< r_s < 6$, and are `strongly coupled' in the sense that $\Gamma\simeq r_s >1$.
\subsection{\bf Exchange-correlation potentials.}
Exchange and correlation potentials used in DFT at $T=0$ are obtained from
quantum Monte Carlo (QMC) calculations for the ground-state energy $E$
of the uniform electron gas.  Finite-$T$ results were available from
 ring-diagram summations (i.e., RPA) from a number of authors
viz.,
 from Rajagopal et al,  Callaway et al., Perrot and Dharma-wardana~\cite{excRPA}. 
 Models beyond  RPA have been
given by Dandrea, Carlsson and Ashcroft,  and
 by Iyetomi and Ichimaru~\cite{IchiIy}. These models fail to
ensure a positive $g(r)$ although this may not seriously affect the
xc-energy calculation. The CHNC avoids the latter problem, with
a non-negative  $g(r)$.
In Fig.~\ref{pairpot.fig} (Right)
we present the finite-$T$ exchange-correlation potential $V_{xc}(r_s,T)$,
 obtained from
the CHNC $g_{ee}(r)$,  in units of the
zero-temperature value of the exchange potential $V_{x}(r_s,T$=0). Recently
QMC-$f_{xc}$  data for finite-$T$ have become available~\cite{Brown13},
 and  agree with previous
results obtained via a coupling-constant integration of $g_{ee}(r)$; the
latter itself has been obtained from various models due to, e.g., 
Iyetomi and Ichimaru,  and the CHNC.
The CHNC generates PDFs of quantum systems via classical maps
 (see sec. ~\ref{eos-H}).
 Sandipan Datta and James Dufty
 have also explored such classical maps, and developed methods that provide
 equivalent results (see Dutta and Dufty~\cite{SPanDufty}).

{\bf Controversial issue.}\\
The results from Hartree-Fock calculations of Sjostrom {\it et al.} for an electron-proton mixture are in qualitative agreement with the parametrized finite-$T$ exchange potential given by Perrot and Dharma-wardana~\cite{Sjostrom1}.

 The xc-functionals derived from the electron subsystem alone are used in solid-state and quantum chemistry calculations. The electrons are in the `external potential' of fixed ions, and such $f_{xc}([n])$
are not fully adapted to WDM calculations where the ion density $\rho(r)$ enters in a direct way. The correlations  involve both $n(r)$ and $\rho(r)$. The ion-correlation potential can be expressed in a fully non-local manner in terms of the classical hyper-netted chain diagrams, as in Eq. 3.4 of Ref.~\cite{hyd0}. However, the correlation-sphere-DFT-HNC method of  Ref.~\cite{hyd0} did not provide a method of obtaining the  $g_{ee}(r)$ although it provided $g_{ii}(r)$ and $g_{ie}(r)$. This short-coming is now resolved via CHNC. Thus the electron $f_{xc}$ in, say, a hydrogen plasma, calculated from the $g_{ee}(r)$ generated from CHNC is a functional of both $n(r)$ and $\rho(r)$, as presented in Fig. 1(c) of  Ref.~\cite{ppots12}.

\section{Equation of State.}
\label{eos-sec}
In this section we illustrate the calculation of the EOS using two methods,
 applied to two interesting cases. (i) The EOS of Aluminum, using 
the correlation-sphere 
approximated using the NPA model to implement DFT
  for electrons. The ion-ion distribution is
 calculated using the modified-HNC equation (MHNC) that includes a model bridge
 function~\cite{RosAsh}. (ii) We consider a hydrogen plasma via a fully
 self-consistent DFT-based correlation-sphere approach, as well as a
 classical-map HNC (viz., CHNC) model where electrons with two types of spins and ions are all treated as classical particles. In this case parallel-spin electrons carry an additional Pauli interaction.

\subsection{\bf The EOS of Aluminum {\it via} NPA-DFT.}
\label{eos-Al}
We assume spherical symmetry, with $\vec{r}$ 
 denoted by $r$. The system is defined by a correlation sphere of
 radius $r=R_c$, large enough  (e.g.,  $R_c\simeq 6r_{ws}$) 
to include the oscillations of the ion-ion PDF (fig.~\ref{jelcav.fig}).
The Kohn-Sham calculation for
the NPA  gives finite-$T$ bound states, scattering phase shifts, densities
of states, continuum lowering etc., while the MHNC and OZ-equations
give the 
$g_{ii}(r)$. They can  be used to calculate the electron free energy, 
$Zf(V,T)$,
 ion-embedding free energy $F_i(V,T)$, and the ion-ion interaction 
free energy $F_{ii}(V,T)$ in a
 transparent manner~\cite{bep, pdw}.  

\begin{figure}
\centering
\includegraphics[width=14 cm]{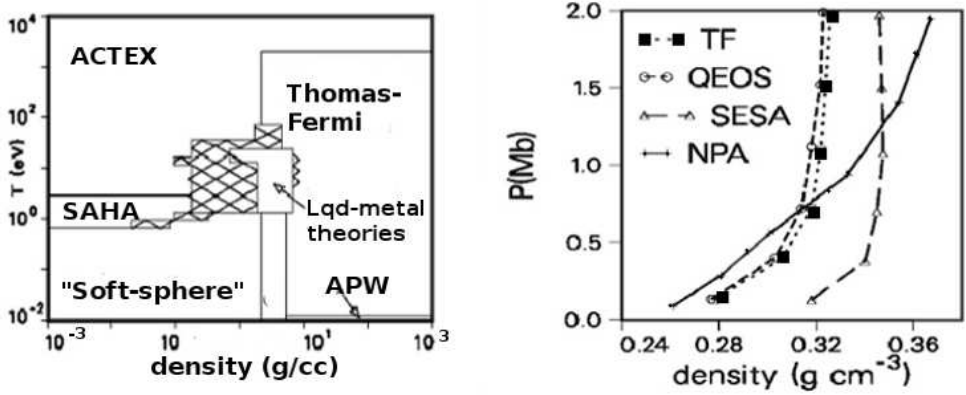}
\caption{(left)
$\rho, T$ plot of the sub-regions used to calculate the SESAME-equation of state
 for aluminum.
 The hatched $\Gamma$(1-50) region is the object of the present NPA calculation, 
to compare with TF, QEOS, and an interpolation from the neighbouring 
regions (SESA). (right ) Hugoniot curves for initial conditions
 $\rho_0$ = 0.05 gcm$^{-3}$ and $k_BT_0$ = 1.0 eV, as
calculated in the four EOS models. As expected QEOS and TF are very similar.
SESAME stands to the right, while the NPA cuts across.}
\label{Al-eos1}
\end{figure}

Fig.~\ref{Al-eos1} (left) shows regions in  $\rho, T$
 and various models used to calculate the EOS. The cross-hatched region 
(1$<\Gamma<50$) is outside the validity of the usual methods, but calculated
 with them here to compare with our NPA results. The resulting Al-Hugoniots are
 for initial conditions $\rho_0$ = 0.05 gcm$^{-3}$ and $T_0$ = 1.0 eV.
 They are given in Fig.~\ref{Al-eos1} (right), for the NPA and for three well-known
 models (for more details, see~\cite{eosb}). 

{\bf Controversial issues.}\par
{\bf (a)} These results show that wide disparities in EOS calculations occur in some `strongly-coupled' parts of the phase diagram,
 even for well-studied `simple' systems like Aluminum.  Furthermore, we have predicted~\cite{pdw} the existence of a plasma-phase transition (PPT) in Aluminum using DFT-NPA methods.  PPTs were first proposed by G. Norman and A. N. Starostin in 1968~\cite{Norman68}. G. Chabrier (for hydrogen, ~1990) and others have discussed PPTs using non-DFT models~\cite{Moscow2005}. There is experimental verifications of PPTs, or perhaps `liquid-liquid' transitions,  but they lack clarity due to problems in the interpretation of the experimental data (e.g., for hydrogen~\cite{Nellis96,Nellis13}) and the lack of good agreement with different theoretical simulations~\cite{LorenzenRedmer10,MoralesCep13}. 

{\bf (b)} The NPA-approach, based on a large correlation sphere is a true DFT
model where the electrons are replaced by {\it non-interacting} Kohn-Sham electrons
at the {\it interacting density}. Hence the chemical potential of the
 electrons is $\mu^0$, while the (space-dependent) corrections  $\mu(r)$-$\mu^0$
constitute the function $V_{xc}(r)$. In contrast, ion-sphere models like the {\it Inferno} and {\it Purgtario}
 do not rigorously map the electron system to a non-interacting system. Hence they
use a chemical potential $\mu\ne\mu^0$ constructed to recover the expected average
electron density $\bar{n}$ within the ion-sphere. Such models, while being quite useful, are probably not true DFT models. 

\subsection{\bf EOS of Hydrogen using a classical map for the electrons.}
\label{eos-H}
Hydrogen is of great interest for
 its technological importance, as a fusion fuel, and in basic physics. Ginzburg regarded hydrogen as a `key problem in Physics and Astrophysics', perhaps noting the key questions of metal-insulator transitions, atomic to molecular transitions, structural phase transitions, quantum melting, the role of nuclear-spin and isotope effects, Fermi-liquid physics etc. 
Thus many presentations  at past SCCS meetings 
 cover various aspects of hydrogen physics 
\cite{HessEbl86, Chihara86, Kalman97, Ashcroft2002}. Hydrogen has also been a popular test case for various statistical potentials proposed to map quantum systems to classical systems.

If the interactions in an electron-ion system (e.g., an electron-proton system)  could be modeled by effective classical interactions containing corrections for quantum effects, then classical simulations of quantum systems can be attempted. This approach has a long history, where $N$-particle wavefunctions are used to define $N$-body potentials via Slater sums~\cite{Kraeft86}. For example, the two-body potentials are given by:
\begin{equation}
S_{ab}(r)=\exp\{-\beta U_{ab}(r)\}= const \sum_j \exp(-\beta E_j)|\Psi_j(r)|^2 .
\end{equation}
Here $\Psi_j$ is the $j$-th eigenstate of the pair $a,b$. The inverse temperature is $\beta$, while $E_j$ is the eigen-energy. The interaction
potential between the pair $a,b$ is $U_{ab}(r)$. Wavefunctions calculated to first order in perturbation theory were used by Kelbg to construct $U$, while simplified forms for $U$ were presented by Deutsch et al~\cite{Deutsch73}. Complementary sets of potentials using momentum states are also needed. Although there is a significant literature on such efforts, the regime of validity is restricted to weakly degenerate, weakly coupled quantum systems. 

The CHNC approach does not use the density matrix, but exploits the
general applicability of the pair-distribution concept to classical or quantum systems, at $T=0$ or finite-$T$. The classical Coulomb fluid with a PDF closely matching  $g_{ee}(r)$ provides our classical map.
This mapping is simplest if there are no bound states.  
Dharma-wardana and Perrot proposed that the
quantum PDFs of the uniform electron gas at a physical temperature $T$
 can be accurately recovered as those of a classical Coulomb fluid  at a 
 temperature $T_{cf}$ such that $T_{cf}=(T_q^2+T^2)^{1/2}$, where $T_q$ 
depends only on $r_s$, the electron Wigner-Seitz radius.
The diffraction-corrected Coulomb interactions were augmented with a 
`Pauli-exclusion' potential $P(r)$ for parallel-spin electrons defined to 
be the potential which gives the parallel-spin non-interacting PDF, 
i.e.,  $g^0_{ee}(r)$ at the given physical temperature. This is a 
generalization (to finite-$T$) of an old idea due to Lado~\cite{Lado67}, 
that had languished since 1967. A trial equilibrium state should reach 
the total energy given by, say, a DFT calculation. Since only the
 many-body energy needs to be matched, the quantum temperature $T_q$ was 
chosen such that the classical Coulomb fluid had the same DFT-correlation 
energy as the electron fluid at $T=0$, known accurately from quantum simulations~\cite{atta}. 

Unlike the earlier methods of Singwi et al~\cite{Mahantxt}, or Ichimaru et al, the $g(r)$ calculated by CHNC is positive for all $r$ and all densities due to its Boltzmann form, and obeys various sum rules by construction. 
The method becomes exact at sufficiently high $T$, and also for non-interacting electrons at all $T$. This CHNC scheme was
 verified to give quite good results, by comparing with QMC $g(r)$ in 2D and 3D for strongly interacting electrons at the extreme quantum limit of $T=0$, where
the coupling constant becomes equivalent to $r_s$. In effect, just a single parameter, $T_q$, was the main input to the CHNC theory to make it work for interacting electrons 
at $T=0$. A tabulation for $T_q$ as a function of $r_s$ was provided for both 3D and 2D electron fluids~\cite{prl1,prl2}.
 Most of these results were independently
 verified for the 3D case at $T=0$, and at finite-$T$ by
 Datta and Dufty~\cite{SPanDufty}. Related work by Bulutay and Tanatar (buluty2002), and by Totsuji (Totsuji04) for the 2D-electron system should be mentioned.

\begin{figure}
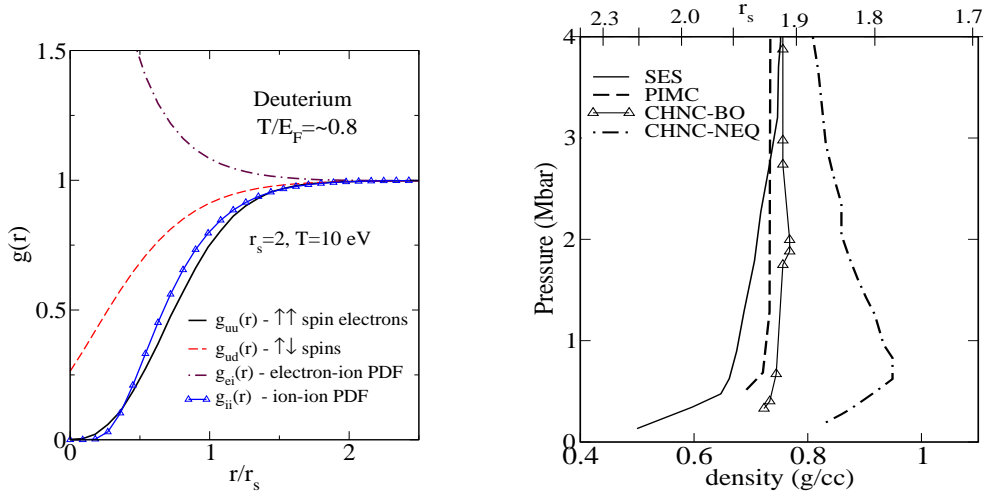

\vspace{0.3 cm}
\centering
\includegraphics[width=55mm, height=60 mm]{grt10rs2.eps}
\hfil
\includegraphics[width=60 mm, height=65 mm]{newHug.eps}
\caption
{(Left) The spin-resolved PDFs $g_{ee}(r), g_{ie}(r)$, and $g_{ii}(r)$ for a
 Deuterium plasma at $r_s$=2 and at $T$=10 eV, calculated {\it via} CHNC. 
As $r\to0$, the 'at-the-nucleus' value of $g_{ei}(r)\to \sim3.9$.
(Right) Comparison of the CHNC with  the SESAME and 
Path-integral Monte Carlo (PIMC) Hugoniots for compressed deuterium.
A non-equilibrium Hugoniot obtained using CHNC is also displayed 
(see Ref.~\cite{chncHyd} for details).
}
\label{fighu}
\end{figure} 

The full correlation-sphere-DFT-HNC treatment of a hydrogen plasma, inclusive of bound-states was given by Dharma-wardana and Perrot in 1982~\cite{hyd0}. The reported $g(r)$ agreed closely with the Hanson-MacDonald simulations.  As already noted, the correlation-sphere-DFT-HNC does not provide a $g_{ee}(r)$, while CHNC does.  
%

{\bf Controversial Issues.}\\
Consider an electron-ion mixture, species $s=e,i$, with deuterium as
the example~\cite{chncHyd}. With $T=T_e=T_i$ as
 the physical temperature, $T_{cf}$ for the `classical-map' electrons, 
the classical MHNC equations and OZ equations
  provide the PDFs $g_{ss'}(r)$.
 The shock Hugoniot
for D obtained by this method agreed closely with the
SESAME tabulations, and with path-integral Monte Carlo (PIMC) results.
A non-equilibrium case, where the temperature associated with the
deuteron-electron interaction in the HNC equation is different from the nominal
temperature is also shown in Fig.~\ref{fighu}. Such
non-equilibrium Hugoniots give softer compressed plasmas, reminding
 the much-debated Laser experiments from the
Livermore laboratory~\cite{HugExp1}. In our view, this topic should be
considered as still open. The Hugoniots traditionally calculated using VASP etc.,
 use the Born-Oppenheimer approximation and are in effect limited to
equilibrium systems.

\section{Static ($\omega=0$) transport properties}
The Kubo formula provides a direct route to the linear transport 
properties of a system if the equilibrium correlation functions and
 pseudopotentials are available. The Ziman formula for the electrical
conductivity~\cite{Mahantxt} of metallic systems can be related to the Kubo formula,
or to the Boltzmann equation, using the concept of an average
collision time $\tau$~\cite{res1,pdw,lvm,PDW-Thermophys}. 
The electrical conductivity can also be related
 to the thermal conductivity using the Wiedemann-Franz law.

\begin{figure}
\centering
\vspace{0.3 cm}
\includegraphics[width=75mm, height=85 mm]{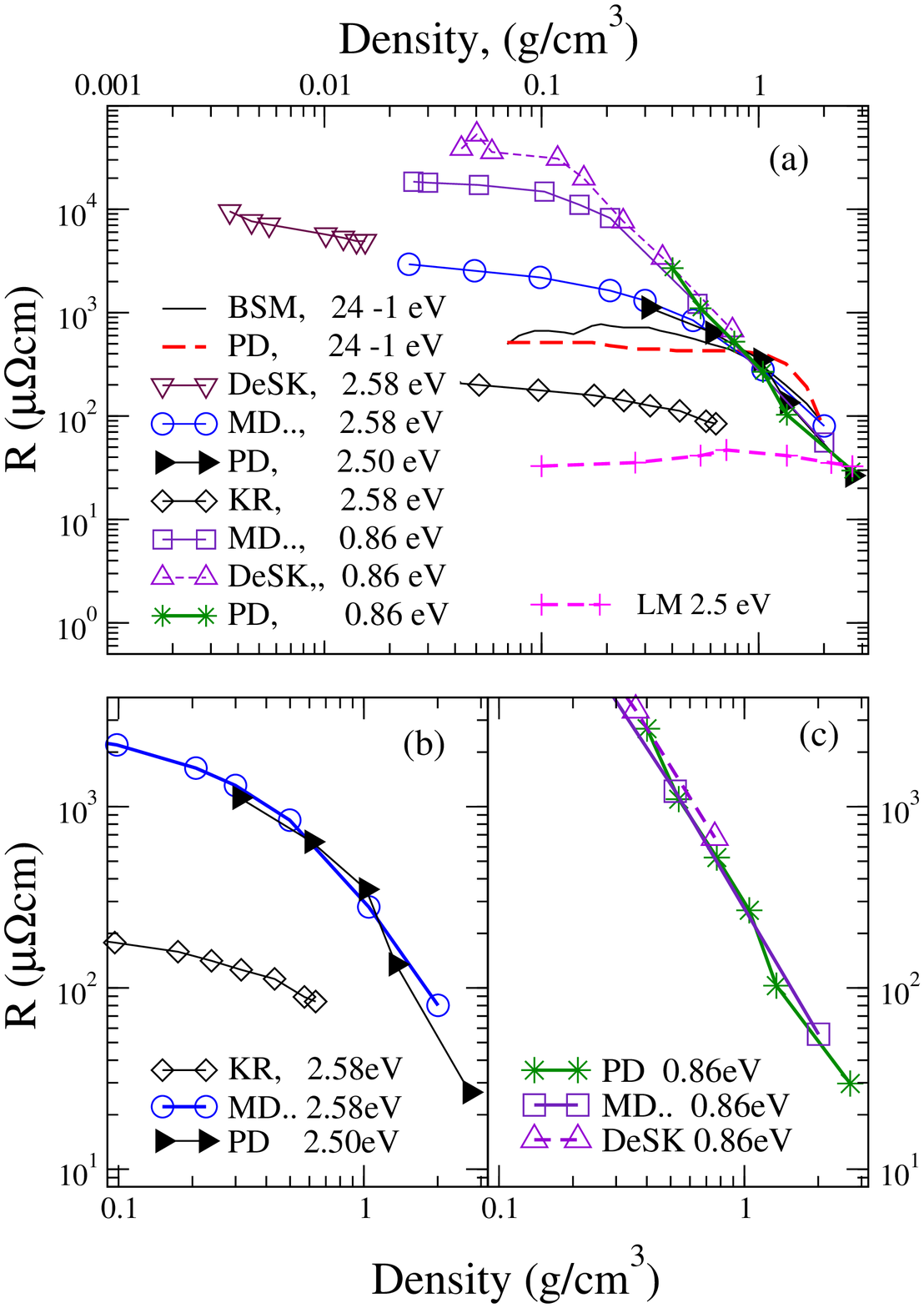}
\hfil
\includegraphics[width=80 mm, height=85 mm]{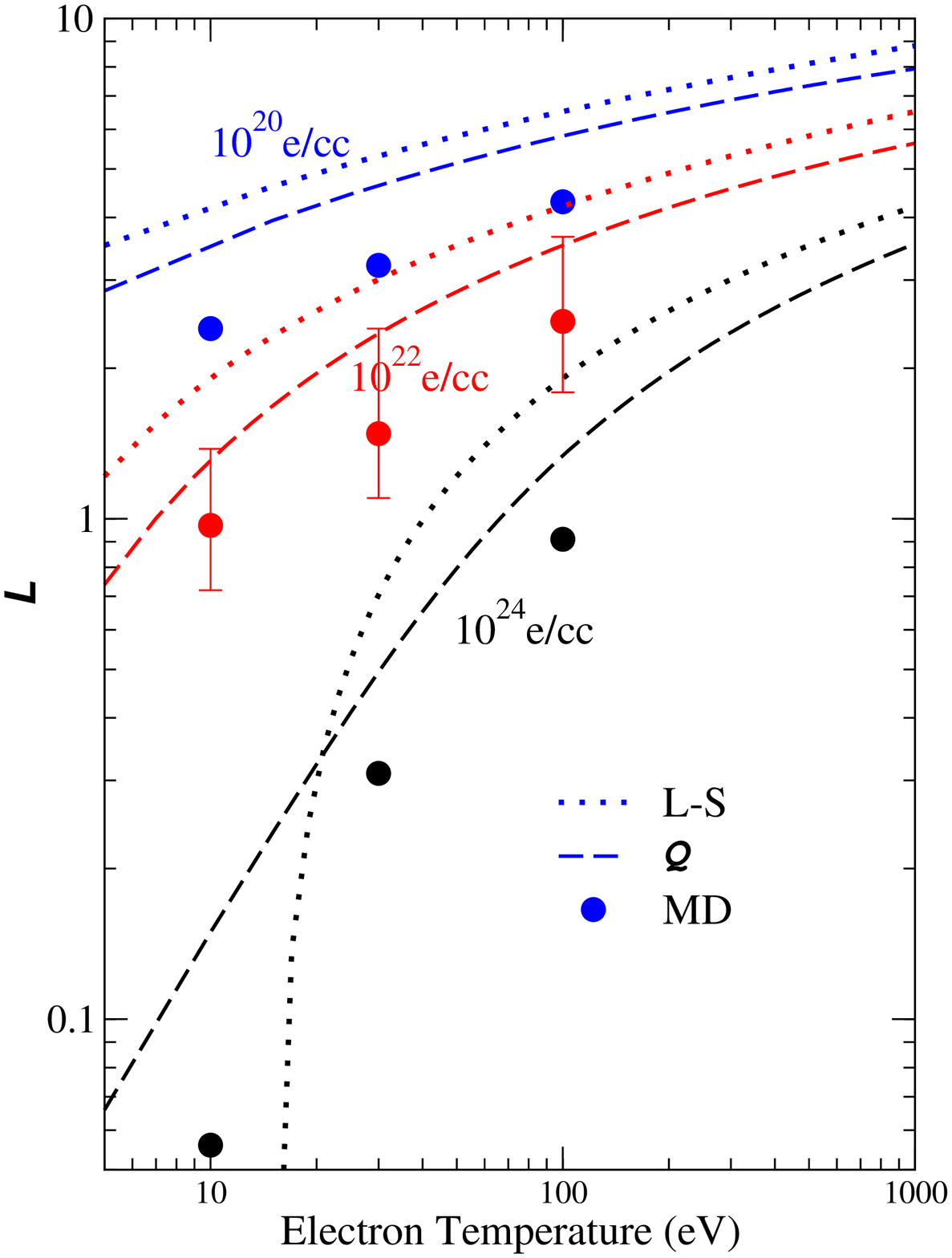}
\caption
{({\bf Left}) {\it Comparison of experimental and theoretical
results for the resistivity of Al plasmas}.  Panel (a): BSM
 - experimental
 results from ref.~\cite{benageR}, Benage {\it et al.,}
 $T$ ranges from $\sim$24 to 1 eV. Theory,
 Perrot and Dharma-wardana, labeled PD, ref.~\cite{pdw,res1} are
 for the BSM range of $T$; also at $T$= 0.86 eV (10$^3$K)
 and 2.5 eV ($\sim3\times 10^3$K). Selected
 experimental data of
 DeSilva {\it et al.,} ref.~\cite{desk} at 10$^3$K and $3\times 10^3$K are
 labeled DeSK. The DFT molecular-dynamics
 results of Mazevet {\it et  al.}~\cite{Mazevet07}, labeled MD,
 are at 10$^3$K and $3\times 10^3$K.
 Results from the ``chemical'' model of Kuhlbrot {\it et al.,}, KR,
 ref.~\cite{kuhl}.
 The Thomas-Fermi model of Lee and
 More (LM), ref.~\cite{leemore} are given only at $\sim3\times 10^3$K.
 Panels (b) and (c) show the excellent agreement between the two
 first-principles calculations, MD and PD, in the warm {\it dense}
 matter range ($\rho> 0.1$g/cc),
 at 10$^3$K and $3\times 10^3$K respectively.
({\bf Right}) {\it  Energy Relaxation}:
The Landau-Spitzer classical Coulomb Logarithm (${\cal L}$) for temperature
relaxation in Hydrogen is compared for three densities ($10^{20},10^{22},10^{24}$ e/cc)
with classical MD results (filled circles)~\cite{elr-boundst} and
with the analytic formula for the Coulomb factor ${\cal Q}$  
inclusive of quantum effects (dashes), given by Dharma-wardana~\cite{elr-boundst},
Error bars~\cite{elr-boundst} are shown ({\it n.b.}  log axis) for the MD data are displayed at one density only.
}
\label{rcomp}
\end{figure}

The Ziman formula requires the ion-ion structure factor $S_{ii}(k)$, the
electron-ion pseudopotential $V_{ie}(k)$ and the dielectric function
 $\epsilon(k)$ for a calculation of the
 resistivity $R$ of a SCCS. Implementations
 using the $S(k)$ from the classical OCP, with $V_{ie}(r)=-Z/r$ 
(a point ion model), and using the simple Thomas-Fermi or Lindhard screening
had been implemented by Claude Deutsch, George Rinker and others.  Structure factors and
$V_{ie}$  obtained from DFT-NPA models were used by
Dharma-wardana, Perrot, and Aers for resistivity calculations using the
Ziman formula and its extensions, as given in their publications~\cite{cdwAers83,pdw,lvm}.
They obtained excellent agreement with the then available resistivity data
for aluminum and other plasmas, and for several liquid metals (Fig.~\ref{rcomp}). Many of these developments were
presented at the SCCS meetings, and in other publications.

\subsection{\bf Controversial issues in conductivity.}
\label{contro-cond}
(a) {\bf Contributions from e-e scattering.} 
We have  emphasized that there are
 {\it no contributions to the
resistivity from electron-electron scattering}  if 
 the electron eigenfunctions
can be regarded as plane-wave~\cite{lvm}. In such cases the
 one-electron current operator
$\vec{j}$ commutes with $H_{ee}$ and hence there is no contribution to 
the resistivity~\cite{lvm}. However, interactions and temperature can 
create localization where the electron wavefunction is not 
a plane wave. Electron dynamics becomes hopping 
motion~\cite{dwp-hopping92}, diffusive motion etc., and a 
different regime (e.g., as in an electron glass, or systems with mobility edges) applies. Electron-
electron scattering enters into the resistivity of periodic structures 
via {\it umklapp} processes, while boundary scattering and band-folding 
effects are important in nanostructures. A different discussion given 
by  Reinholz et al.,~\cite{rein-rop14} should also be perused.

(b) {\bf Minimum-conductivity plateaus.} The first experiments on the short-time two-$T$ 
conductivity $\sigma(T_i,T_e)$ of laser-shocked materials already measured the  dynamic
conductivity $\sigma(\omega)$ (of laser-pulse heated aluminum) up to $T_e \sim$ 100 eV, along the Shock
Hugoniot up to 20 Mbars~\cite{R-Milsch-Alu92}, and reported $\omega \to 0$ data for $\sigma$. The data show three stages as a function 
of $T_e$ while $T_i$ remains constant (see Sec.~\ref{elr}): (i) $\sigma$ 
decreases  with the electron temperature $T_e$, (ii) $\sigma$ reaches  a 
`minimum-conductivity'  plateau (MCP) and then (iii) begins to  rapidly 
increase with $T_e$. A number of authors,  borrowing N. 
F. Mott's ideas, have claimed that the MCP arises when the mean free-
path $\lambda$ becomes comparable to the mean inter-ionic distance. Unlike in semiconductors, this 
argument is manifestly incorrect in metallic WDM systems. Here the MCP 
occurs when the chemical potential $\mu$ passes through zero and becomes 
negative. Before the MCP, only the electrons  within a short window $\pm 
T_e$ near $E_F$ contribute to $\sigma$ (n.b., $k_B=1$ in our units). 
After the MCP, $\mu$ is negative, and the electrons become non-degenerate. 
The whole Fermi sphere is rapidly destroyed and all electrons (not just 
those near $E_F$) contribute. Then $\sigma$ rapidly increases as in 
Spitzer theory. This theory can quantitatively account for the experimental observations~\cite{DynRes92}.

\section{Plasma spectroscopy: dynamic conductivity and opacity.}
\label{opacity}
The linear optical absorption by a strongly-coupled Coulomb system inclusive
 of free-free, bound-free and bound-bound processes can be written in terms of
 the dipole-dipole correlation function of the system, or in terms of current-current
correlation functions for bremstrahlung like processes.
The absorption cross-section is related to the $S(q,\omega)$ of the electron subsystem, since it is the electrons that respond to the external field.

 It is more commonly (and equivalently) treated in terms of 
the complex dielectric function $\epsilon(\omega)$ or the complex conductivity $\sigma(\omega)$ of the system. The distinction between transverse and longitudinal dielectric
functions becomes unimportant since the photon-momentum $q$ is small compared to other momenta in bulk matter
(but not so in nano-structures). Nevertheless, the careful discussion needs to include the $\vec{q}$-dependence.

An important issue is the modification of the external radiation field $\phi_{ext}(\vec{r},\omega)$ inside the plasma due to induced fields, so that the
effective field $\phi_{eff}(\vec{r},\omega)$ becomes complex and time lagged (i.e., phase shifted) due to the density displacement $n_{ind}(\vec{r},\omega)$ induced in the system.
\begin{equation}
\label{zw.eqn}
n(\vec{r})=n^0(\vec{r})+n_{ind}(\vec{r},\omega),\;\;n^0=\sum_m|\psi^{ks}_m(\vec{r})|^2,\;\; n_{ind}=
\int\chi(\vec{r'},\vec{r})\phi(\vec{r'},\omega)d\vec{r'} 
\end{equation} 
Here the induced charge is treated in linear Response using $\chi(\vec{r'},\vec{r})$. The induced charge produces induced Coulomb fields and corrections to exchange and correlation.
These modify the poles of the response functions to totally change the `single-particle' features of the spectrum, in agreement with observed photo-ionization spectra.

On the other-hand, the commonly used form of the Kubo-Greenwood formula assumes a `non-interacting electron' model (i.e, uses `single-electron' states $|n>$  with energies $\epsilon_n$ and
occupation numbers $f_n$, like in Hartree-Fock theory, or Kohn-Sham theory).
Then the complex conductivity tensor (atomic units) is:
\begin{equation}
\label{Green-Kubo.eqn}
\sigma_{\alpha,\beta}(\omega)=-i\sum_{m,n}\{f_m-f_n\}
\frac{<n|j_{\alpha}|m><m|j_{\beta}|n>}
{(\omega+i\eta)+\epsilon_m-\epsilon_n}
\end{equation}
The $j_{\alpha}$ is the $\alpha$ = $x,y$ or $z$ component of the current $\vec{j}$=$-(|e|/m_e)\sum_i\vec{p}_i$ , with $p_i$ the momentum of the $i$-th electron of mass $m_e$ and charge -$|e|$ summed over electrons per unit volume, while $\eta=1/\tau$ is an infinitesimal in a perfect crystal. For the diagonal case $m$=$n$, the label $n$=$\vec{k},b,s$ becomes the momentum state $k$ with the band index $b$ and spin index $s$, while $m$=$\vec{k}+\vec{q},b,s$, where $\vec{q}$ $\sim0$ is the photon momentum, with $\tau$ becoming a momentum relaxation time. Thus the diagonal term  $(f_m-f_n)/(\omega+\epsilon_m-\epsilon_n), m=n, \omega\to 0 $ is simply $\partial (f(\epsilon_{\vec{k}b}))/\partial\epsilon_{\vec{k}b}$. This leads to the usual {\it intraband}  Drude conductivity, while the $m\ne n$ terms provide the inter-band contributions. 

The use of Eq.~\ref{Green-Kubo.eqn} does not give agreement with observed photoionization spectra of atoms. The inclusion of the induced terms, viz., Eq.~\ref{zw.eqn} is essential in matching observations.
The first successful calculations in atomic physics of these effects were given by Zangwill and Soven at $T=0$. Applications to systems with delocalized electrons (e.g., WDM) is more challenging due to the presence of continuum states.  Grimaldi, Lacourte and Dharma-wardana presented such a calculation for an iron plasma at 100 eV, considering bound-bound, bound-free and free-free processes, and reported the calculations at the Santa Cruz SCCS-meeting as long ago as 1986~\cite{Grimaldi, SantaCruz}. The effect of such processes on plasma opacities, radiative transfer, etc., remains open as these calculations have not been followed up, although this has now become the domain of time-dependent DFT. These effects are, however, known to be not important for the $K$-edge~\cite{kedge}.

{\bf Controversial issues in plasma spectroscopy.}\par
{\bf (a)} The  ion microfield can be expressed in terms of the ion-ion 
PDF, or by a re-summation process, e.g., as in Baranger and Mozer,
 or more recently as in Iglesias {\it et  al} (see Review~\cite{Demura2009}).
 However,  although the ion-ion interaction is treated by partial summation to ``all orders",
 the ion-electron interaction is commonly treated
 linearly, and classically, using Debye screening for a point-ion,
 even for very high-$Z$ ions!  This can be avoided by DFT-NPA methods.
 Furitani, Perrot and Dharma-wardana presented calculations of ion microfields where the full quantum electron-ion interaction was used
to obtain $n(r)$; furthermore, an alternative method of re-summation to all
 orders of the microfield was also given~\cite{Furitani}. Unfortunately,
Debye-model based microfields are still in vogue.

{\bf (b)} The center of spectral lines is affected by ion-dynamics.
While molecular-dynamics simulations can be used to study the effect, transparent analytical treatments are usually not available. In our view, the simplest treatment of this is via the coupled-mode effect~\cite{elr2001,elr0}. The ion subsystem and the electron subsystem cannot be treated as independent subsystems for sufficiently long time scales (i.e, small $\Delta\omega$, near the center of an absorption line). In such regimes, the ion-plasma oscillations have time to become ion-acoustic excitations. This process also has dramatic effects in 
slowing down energy relaxation rates of two-$T$ quasi-equilibrium plasmas, as described below.

\section{Laser- and shock- created plasmas, their energy 
relaxation (ER) and transient  properties.}
\label{elr}
Femto-second lasers acting on a SCCS can deposit energy into
 the electrons faster than the rate of relaxation of the deposited energy
 into the ion subsystem. Thus two-temperature SCCS in quasi-equilibrium 
for short-time scales $t$ such that $\tau_{ei}>t>\tau_{ee}$, can be created. 
Here $\tau_{ss'}$ is the temperature relaxation time between species $s,s'$.
 Normally, $\tau_{ee}$ is of the order of femto-seconds, while
 $\tau_{ei} > \tau_{ii}$ is in the pico-second range. 
Alternatively, the ions can be preferentially heated to create SCCS with $T_i>T_e$,
using mechanical shocks.

 Andrew Ng  found that short-pulse laser-generated Si-plasmas ($T_e > T_i$) showed equilibration rates significantly slower than those given by standard formulations. Similar issues had arisen in the carrier relaxation of hot semi-conductors via longitudinal optical (LO) phonons. The common consensus in the plasma-physics community was that standard Landau-Spitzer (LS) formulations had been largely confirmed by, e.g., the molecular dynamics simulations of Hanson and MacDonald~\cite{HM-md}. These models were based on classical-trajectory theories using `Coulomb Logarithms'  to regularize collision integrals, as in LS theory, and sometimes incorporating  intuitive  `minimum free-path' concepts.

However, {\it ad hoc} models can be avoided using  realistic screened electron-ion potentials and structure factors from DFT. Dharma-wardana and Perrot used the Fermi Golden Rule (FGR) as well as {\it the assumption} of the existence of fluctuation-dissipation theorems (FDT) applicable to each subsystem (ion subsystem and the electron subsystem) since they had identifiable temperatures $T_e$ and $T_i$. Since this was a non-equilibrium system, the results were checked using the Martin-Schwinger-Keldysh non-equilibrium Green's function technique to calculate ER rates which are directly related to temperature relaxation times~\cite{elr0}.  

Numerical calculations showed that the FGR+FDT results already gave relaxation rates slower than those of LS. Furthermore, for times $t$ long enough to allow the formation of ion-acoustic modes by the electron screening of bare-ion plasmon modes, with $t<\tau_{ie}$, the ER occurs via coupled-mode processes (ion-acoustic modes). These were found to sharply lower the rate of relaxation of hot plasmas. The results were in much better agreement with the experiments of Ng {\it et al.}, and the Belfast~\cite{Belfast} experiments. 

The use of the $f$-sum rule to evaluate the FGR+FDT formula was, in our view, a significant step in obtaining convenient forms for energy relaxation. This avoids the difficult problem of modeling the $\omega$-dependent local field factors contained in the response functions. The theory can be applied to SCCS like molten aluminum~\cite{cdwELR01}.

 Nearly closed-form  analytic formulae for energy relaxation, inclusive of important quantum effects have been given by 
Dharma-wardana~\cite{elr-boundst}, and by  Brown {\it et  al.} (BJPS)~\cite{BJPS}. The BJPS formula, which delineates cutoffs by dimensional regularization, and the formula by Dharma-wardana, viz., Eq. (13) in Ref.~\cite{elr-boundst} containing physically regularized integrals, look quite different, and 
were derived from entirely different considerations;  but they turn out to be numerically quite close. The theory of ER from bound states is presented for a Hydrogen plasma in Ref.~\cite{elr-boundst}.  The analytical-$f$-sum formula of Dharma-wardana (without bound-state 
effects) is compared with MD simulation results by Glosli et al~\cite{mmll}  and LS in Fig.~\ref{rcomp} (right panel). The MD results are lower than Dharma-wardana's analytical $f$-sum results, and may include coupled-mode effects. 

Meanwhile, the theoretical effort has been resumed from the approach of quantum kinetic equations (QKE) by a number of workers including Schlanges, Murillo, Benedict, Daligault, Bornath, Vorberger,  Gericke, Kremp, Kreft and others~\cite{SMVG2010}. It is not possible to examine this  body of work in detail here, except to note that these authors have 
confirmed within QKE many of the results obtained by us, and provided derivations by methods more familiar to the plasma community. Furthermore, a number of new codes for large-scale implementations have been reported~\cite{DD08, Bene12}. Here we make some pertinent observations, mainly in regard to theoretical consistency.

The interest in quantum-kinetic equations, e.g., the Lenard-Balescu equation (LBE), resides partly in that they may be used directly in simulations, and partly as a theoretical tool. However, the collision integral used in the LBE is in practice nothing but the Fermi Golden rule, and  extensions of LBE are usually equivalent to $T$-matrix formulations and elaborations of the dielectric functions, but lacking in the controls that are available with diagram techniques where 
intermediate-frequency integrations are left intact (unlike in QKE). The same difficulty arises in the Zubarev two-time Green's function approach~\cite{ElrPRL91}, where truncations of the chain of equations of motion are needed. Many of the published papers use generalized-RPA like response functions while  Debye-H\"{u}kel approximations, Yukawa approximations, and assumptions of point ions without a core etc., are hidden in some intermediate steps, and hence consistency is difficult to establish.  

Singwi, Ichimaru and others formulated local-field corrections (LFCs) to the response functions via an $S(k)$ derived from a $g(r)$~\cite{Mahantxt}. This procedure rigorously needs an $S(k,\omega)$. A pitfall here is that important constraints like the compressibility sum-rule is 
violated when, say, HNC generated $g(r)$ are used. Nevertheless some authors have invested considerable efforts in such LFC-construction, when the use of a simple compressibility ratio is perhaps sufficient (see remarks below Eq.~\ref{pseud-pair.eq}). The construction of the frequency-dependent LFC is even more demanding, and needs to satisfy known constrains given in, e.g., sections 5.6-5.9 of Ref.~\cite{GiuVig05}.

Another pitfall is the belief that molecular-dynamics can lead to `bench-mark results' for judging the quality of QKE results or generalized Landau-Spitzer results. If MD-simulations are done with the Deutsch potential, or the Kelb potential, why trust the results in regimes beyond the weakly degenerate systems envisaged by such statistical potentials? We found from our work on the CHNC that in many situations the Slater-sum approach can be profitably replaced by one where the diffraction wavevector $k_{df}$ occurring in the potential, e.g., $(-Z/r)\{1-\exp(-k_{df}r)\}$  should ensure that the electron density $n(r), r\to 0$ resulting from such a potential should agree with that calculated from a DFT-NPA model for that ion. 

The physical content of most implementations, including that in Ref.~\cite{Bene12} probably does not go beyond the FGR+FDT model, but this may be adequate for their objectives. The resulting equations (e.g., their Eq. 20) were not  simplified further using the $f$-sum rule, claiming a need  to retain the $\omega$ dependence due to there being a variety of ions. The $f$-sum rule assumptions are valid to second-order in the density fluctuations (due to the antisymmetry of the relevant piece of the response functions). The errors in (i) the $\omega$ integrations, (ii) in modeling $\omega$-dependent LFCs for ions etc., probably outweigh any claimed advantage over not using the $f$-sum rule.

It may be  noted that the calculation of the local-field correction used in Ref.~\cite{Bene12} (their Eqs. 25-29) can be done to satisfy sum rules, and to include the quantum case automatically, as in Eq. 16 of Ref.~\cite{prl1}, without having to do a Kramers-Kronig evaluation or restricting oneself to the classical limit. 

{\bf Controversial issues.}\\
(a)  Ion-acoustic modes in plasmas, and coupled longitudinal optical phonon-plasmon modes in semiconductor interband excitations~\cite{ElrPRL91, ELR-ssc}, have been invoked in coupled-mode energy relaxation.
There is some confusion in the literature about `when' coupled modes (CM) become relevant in energy-relation. If CMs are to be included, then the $\omega$ integration cannot be done using the standard $f$-sum rule.
 In a non-equilibrium system  CMs become relevant if their time scales for mode formation are short enough for them to appear within the time scale $\tau_{ie}$. On the other hand, when $T_e \gg T_i$, relaxation is very fast and coupled-modes (ion-acoustic excitations) may not have time to form, and relaxation via CM becomes increasingly small.  Such issues should be automatically taken into account through the imaginary parts of green's functions describing the CM. However, RPA-type  approximations fail to capture the necessary details, and give results that could be misleading. Here it should be noted that the Lenard-Balescu Equations are simply the kinetic equations that correspond to the RPA.

(b)  No adequate calculations for laser-pulsed two-T systems inclusive of interband scattering processes (e.g., transition metals) exist. Attempting to extract physical quantities from the  $\sigma(\omega)$  using the Drude formula, while ignoring interband terms (e.g.,  $d$-band transitions in Cu, Ag, or Au), 
is highly questionable, even if used to merely report data~\cite{Ping-Ng06}. The real part of the dynamic susceptibility 
$\varepsilon(\omega)$ 
is influenced by interband absorption even at $\omega$ much smaller than the absorption threshold, due to Karmers-Kronig relations. However, the static conductivity can be extracted from the $\omega \to 0$ limit of $\sigma(\omega)$ as this uses only the imaginary part
of $\varepsilon(\omega)$. 

\section{Search for simpler computational schemes.}
\label{simpler.sec}
The previous sections show that predicting physical properties
of SCCS involves obtaining self-consistent solutions of many-electron, multi-ion problems
under non-equilibrium conditions, at many time steps. Hence a search for simplified but valid  physical models continues in many fronts:\\
(1) Replacement of the many-particle quantum many-body problem by equivalent one-body problems. The most important step here has been the advent of DFT.\\
(2) Possibility of eliminating the Kohn-Sham equation and directly implementing the Hohenberg-Kohn theorem by replacing the kinetic energy operator by a functional $T[n]$ that improves the Thomas-Fermi model~\cite{PerrotKE, TrickeyKE}. The non-locality of the kinetic-energy operator has so far prevented a satisfactory resolution of this problem, although several approximate `orbital-free' approaches (that fail to recover shell-structure) have been advanced.\\
(3) Mapping of quantum systems to classical systems. An example of such an approach is the CHNC, where the issue of the kinetic-energy functional is side-stepped by the use of a ``quantum temperature" $T_q(r_s)$ (
see Sec.~\ref{eos-sec}). \\
(4) The method of effective `statistical' potentials via Slater sums is another semi-classical approach~\cite{Kraeft86}.\par
The  NPA-DFT model provided simple, local, linearized pseudopotentials for $V_{ie}(q)$, and pair-potentials $V_{ii}(q)$ as already presented in our discussions of the NPA-model. These can be used in MD simulations. Pair-potentials alone may not be adequate for SCCS, given the experience from semi-conductors where three-body and four-body terms are important in simulating Si or C. Our experience is that if the system contains free electrons (as in a plasma or a metal), then  the pair-potential approach is generally sufficient, even for some transition metals like gold (here the issues are admittedly more delicate).

The DFT-NPA potentials use linear response. If the electron subsystem is strongly coupled, the CHNC provides a treatment that goes beyond the generalized RPA used in the pair-potentials of Eq.~\ref{pseud-pair.eq}.

{\bf Some implementations of CHNC to electrons of different flavours.}\\
The CHNC method was applied to a study of the phase diagram of two-valley 2D-electron (2V-2D) fluids in Si-MOSFETS using a $T_q$ based the 2D-$E_c$ of Tanatar and Ceperley. Interestingly, it was found that the system became critical and showed phase transitions in agreement with the experimental data of Sashkin {\it et  al.}~\cite{Sash2003,2v2d}. Attaccalite {\it et al}., had reported improved $E_c$ for the one-valley 2D system and found a spin phase transition at $r_s\sim27$. However, when the $T_q$ for CHNC was determined from the improved  2D-$E_c$ of Attaccalite {\it et al.,}  the  criticality in the 2V-2D was not obtained. Direct QMC simulations for the 2V-2D by de Palo and Sanatore also showed no criticality~\cite{dePaloSan}.  Interestingly, further improvement in QMC techniques (e.g., use of twisted boundary conditions) removed the spin transition in the single-layer 2D system as well~\cite{Drum2009}, showing that neither QMC, nor methods using QMC-derived $E_c$ are at present capable of reliable predictions of phase transitions at very low densities and very strong coupling, while experimental results are very sensitive to traces of impurities.

The fractional quantum Hall effect (FQHE) in graphene involves electrons of four different flavours. Thus good quantum simulations are prohibitive. Hence it is a good candidate for CHNC methods.
 
 \subsection{\bf Laughlin's classical-map for the FQHE; application to Graphene.}
 \label{fqhe-sec} 
The FQHE is a unique many-body problem  unraveled  by Laughlin who guessed an approximate  many-body wavefunction for $N$-electrons in a magnetic field in the lowest Landau level of a 2D electron gas~\cite{Laughlin}, after examining the wavefunctions of systems with a few electrons. Laughlin's wavefunction $\Psi(z_1,..z_N)$ for such a system, where $z_i$ are coordinates in the 2D electron sheet (complex numbers), has a product of factor $(z_i-z_j)^m$ for each pair of electrons, with $m$ an odd integer. These multiply an exponential factor. Since $m$ is odd, $\psi$ is antisymmetric. Laughlin treats the density $|\Psi|^2 \mapsto \exp(-\beta V)$  to arise from a classical potential $V$ with an effective temperature $T_{eff}=1/2$ such that the coupling constant $\Gamma$ is $2m$. The electron-electron PDFs and ground state energies of FQHE fluids were calculated from classical fluids with pair interactions of the form -$\Gamma\log(|z_i-z_j|)$. Very interestingly, exact solutions to the case $m=1$ (filled Landau level) were already known from the work of Jancovici who had been a pioneer of SCCS meetings~\cite{JancoviciFQHE}.

Laughlin's wavefunction (spin-polarized electrons in a 2D-layer, i.e., of one flavor) was generalized to several interacting 2D layers by Halperin (see the review~\cite{Gorbig11}). The `in-layer' FQHE correlations for the layer (or flavour) $f$ are controlled by an exponent $m_f$ (odd integer), while inter-layer correlation-exponents between layers $f,g$ denoted by $n_{fg}$ may be any positive integer. If $n_{fg}$=1 they are normal Coulomb interactions, while $n_{fg}=0$ implies  no correlations or interactions. We denote the 3-layer Halperin-Laughlin wavefunction as $\psi(m_1,m_2,m_3:n_1,n_2,n_3)$, with the $n$-values denoting the upper triangle $n_{fg}$ in a 3$\times$3 matrix where $m_f$ are diagonal elements.  
 
Laughlin's HNC-map was generalized to the Haldane-hierarchy of FQHE fractions by MacDonald, Aers and Dharm-wardana~\cite{fqheHNC1}. The method has also been used very recently by Dharma-wardana and Aers~\cite{CDW-Aers} to 
calculate the six $g_{ee}(r)$ of graphene-FQHE system~\cite{Gorbig11} of 
fluids with SU(4) symmetry. The electrons can be of 4 different flavours $f$. The FQHE wavefunction in the lowest Landau level  is of the form $\psi(m_1,\cdots,m_4:n_1,\cdots,n_6)$ where there can be six different $n_{gf}$ values, with $g>f$.

\begin{figure}
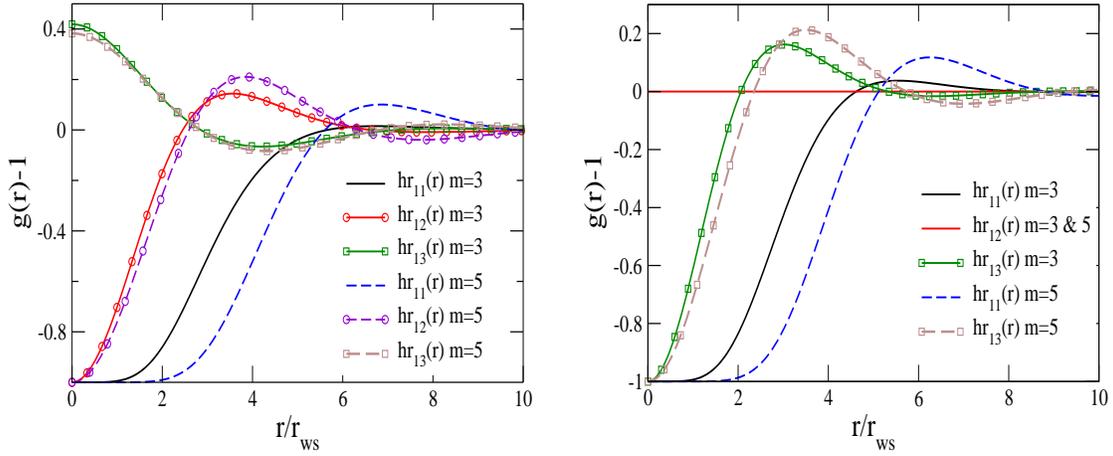

\vspace{0.3 cm}
\centering
\includegraphics[width=70mm,height=60 mm]{hr4-figI.eps}
\hfil
\includegraphics[width=70mm,height=60 mm]{hr4-figII.eps}
\caption
{(Left) The PDFs of a graphene-FQHE system with in-layer exponents $m_f$ all equal
to $m$=3 or 5, while the nearest-neighbour  intra-layer exponents (e.g., $n_{12}$) are unity,
with others (e.g., $n_{13}$) zero. 
These are stable fluids with some on-site (e.g., $g_{13}(r)$) correlations
strongly enhanced.
(Right) Here there is no nearest-neighbour coupling ($g_{12}(r)-1=0$), and the fluids are
unstable (ground state energy $>0$) 
}
\label{fqhe}
\end{figure}

The density $|\Psi|^2$ for the FQHE in graphene, if assumed to be of the 
form  $\bar{n}\exp\{-\beta V(r)\}$ can be regarded as resulting from  four classical fluids 
interacting via $\log|z_i-z_j|$-type potentials, with the PDFs given by 
six-coupled HNC-equations. Dharma-wardana and Aers have solved these 
equations in selected landscapes of $\{m_f, n_{fg}\}$. 
In Fig.~\ref{fqhe} we show the PDFs for several cases. Thus with $m_f=3$ 
or 5 and simple Coulomb interactions between flavours, $n_{fg}=1$, restricted to nearest-neighbour interactions, we get stable Laughlin-
like FQHE in each layer, as well as the phenomenon of enhanced `on-top-
of each-other' type density correlations between next-nearest-neighbour 
layers. Such interlayer correlations have become topical in related fileds of study as well~\cite{Nielson-Perali-13}. Almost all the properties of the FQHE system can be calculated from 
such PDFs. Novel results for multi-layer FQHE systems as well as graphene 
obtained using  the classical map will be presented in detail elsewhere.

\section{Conclusion}
The interest in strongly-coupled Coulomb systems displayed in the early SCCS meetings  began at a time when Zubarev or Keldysh Green functions, BBGKY-type kinetic equations and other many-body methods had come into the field and seemed to provide a spring board to the future. However, while they provided much theoretical and physical insight, the definitive predictive capability  came with the advent of density functional theory, coupled with advances in numerical-simulation techniques like molecular dynamics and quantum Monte Carlo. In our view, even with available computers, the further simplification of the quantum problems via classical maps, development of accurate  kinetic-energy functionals etc., may lead to an order of magnitude simplification of the theoretical calculations. These  provide the inputs to more macroscopic codes. The classical maps are particularly appropriate for use in such simulation codes and open up means of investigating complex systems like FQHE on graphene, which are at present well beyond the reach of quantum simulations..



\begin{thebibliography}{99}
\bibitem{Graziani11}
F. R. Graziani {\it et  al.} Lawrence Livermore National Laboratory
report, USA,  LLNL-JRNL-469771 (2011)

\bibitem{MuriDW2013}
Michael S. Murillo, Jon Weisheit, Stephanie B. Hansen, and 
M. W. C. Dharma-wardana. Phys. Rev. E {\bf 87}, 063113 (2013)

\bibitem{Kalman77}
 Gabor Kalman and Paul Carini, (Eds.)  Strongly Coupled Plasmas (Plenum, New York 1978)

\bibitem{Moscow2005}
V. E. Fortov, K. I. Golden, and G. E. Norman (Eds.),
 J. Phys A: {\bf 39}, 441-4479 (2006)

\bibitem{OZ14}
L. S. Ornstein and F. Zernike,  Proc. Acad. Sci. Amsterdam  {\bf 17}, 793-806 (1914)

\bibitem{DFT-source} W. Kohn and L.J. Sham, Phys. Rev. {\bf 140}, A1133 (1965).
N. D. Mermin, Phys. Rev. A {\bf 137} 1441 (1965)

\bibitem{ZangDFT14}
A. Zangwill, arXiv:1403.5164v1 [physics.hist-ph] (2014)

\bibitem{excRPA}
F. Perrot and M.W.C. Dharma-wardana,  Phys. Rev. A {\bf 30}, 2619
(1984).

\bibitem{prl1}
M. W. C. Dharma-wardana and F. Perrot, Phys. Rev. Lett. {\bf 84}, 959 (2000)


\bibitem{chncPRB}
F. Perrot and M. W. C. Dharma-wardana, Phys. Rev. B {\bf 62}, 16536 (2000);
{\it Irratum} {\bf 67}, 79901 (2003)

\bibitem{Brown13}
Ethan W. Brown, Jonathan L. DuBois, Markus Holzman, David M. Ceperley, Phys. Rev.  B {\bf 88}, 081102(R) (2013) 

\bibitem{drt} L. Dagens, M. Rasolt, and R. Taylor, Phys. Rev. B. {\bf 11}, 2726 (1975)

\bibitem{bep}
F. Perrot, Phys. Rev. B {\bf 47},  570 (1993)

\bibitem{ppots12}
M. W. C. Dharma-wardana,
Phys. Rev. E {\bf 86}, 036407 (2012)

\bibitem{hyd0}
M.W.C. Dharma-wardana and F. Perrot, Phys. Rev. A {\bf 26}, 2096 (1982)


\bibitem{SantaCruz}
M.W.C. Dharma-wardana, Density Functional Methods in Hot Dense Plasmas
in  Strongly Coupled Plasma Physics, (Eds.) Hugh                               
DeWitt and F. Rogers, NATO Advanced Study Series, Santa Cruz.  
 (Plenum, New York  1987) p. 275-293.  
\bibitem{pdw}
 F. Perrot and M. W. C. Dharma-wardana,  Phys. Rev. E {\bf 52}, 5352 (1995)

\bibitem{res1}
Francois Perrot and M.W.C. Dharma-wardana,  Phys. Rev. A {\bf 36}, 238 (1987)

\bibitem{lvm}
M. W. C. Dharma-wardana,  Phys. Rev. E {\bf 73}, 036401 (2006)

\bibitem{Grimaldi}
F. Grimaldi, A. Grimaldi-Lecourt, and M.W.C. Dharma-wardana, Phys. Rev. A  {\bf 32}, 1063 (1985)

\bibitem{elr0}
M. W. C. Dharma-wardana and F. Perrot,  Phys. Rev. E {\bf 58}, 3705, (1998); {\it Errratum} Phys. Rev. E {\bf 63}, 069901 (2000).

\bibitem{hazak}
 G. Hazak, Z. Zinamon, Y. Rosenfeld, M. W. C. Dharma-wardana, Phys. Rev. E {\bf 64},
066411 (2001)

\bibitem{elr2001}
M. W. C. Dharma-wardana, Phys. Rev. E {\bf 64}, 035432(R) (2001)

\bibitem{elr-boundst}
M. W. C. Dharma-wardana,  Phys. Rev. Lett. {\bf 101}, 035002 (2008)

\bibitem{chncPRL3d}
M. W. C. Dharma-wardana, and F. Perrot, Phys. Rev. Lett.,
{\bf 84}, 959 (2000)

\bibitem{chncPRL2d}
Francois Perrot and M. W. C. Dharma-wardana,
  Phys. Rev. Lett., {\bf 87}, 206404 (2001)

\bibitem{chncHyd}
M. W. C. Dharma-wardana and F. Perrot, 
 Phys. Rev. B {\bf 66}, 014110 (2002)

\bibitem{fqheHNC1}
A. H. MacDonald, G. C. Aers and M. W. C. Dharma-wardana, Phys. Rev. B {\bf 31}, 5529 (1985).

\bibitem{cdwPisa2013}
M. W. C. Dharma-wardana,
J. Phys. Conf. Ser. i{\bf 442}, 012030 (2013)
doi:10.1088/1742-6596/442/1/012030


\bibitem{apvmm2013}
Chandre Dharma-wardana, A physicist's view of matter and mind, Ch.8-9 (World Scientific, Singapore 2013).

\bibitem{MurDW2013}
Michael S. Murillo, Jon Weisheit, Stephanie B. Hansen, and M. W. C. Dharma-wardana, Phys. Rev. E  {\bf 87}, 063113 (2013).

\bibitem{Drum2009}
N. D. Drummond and R. J. Needs Phys. Rev. B {\bf 79}, 085414 (2009).

\bibitem{atta}
C. Attaccalite, S. Moroni, P. Gori-Giorgi, and G. B. Bachelet, Phys. Rev. Lett. {\bf 88}, 256601 (2002͒).

\bibitem{Dagens73}
L. Dagens, J. de Physique, {\bf 34}, 8 (1973); J. Phys. C {\bf 5}, 2333 (1972).

\bibitem{Trickey}
V. V. Karasiev, T. Sjostrom, S. B. Trickey, Phys. Rev. E. {\bf 86}, 056704 (2012).

\bibitem{KhannaGlyde76}
 F.C. Khanna and H.R. Glyde. Can. J. Phys. {\bf 54}, 648 (1976); 
C. Gouedard and C. Deutsch, J. Math. Phys. {\bf 19}, 32  (1978).

\bibitem{cdwAers83}
M.W.C. Dharma-wardana and G.C. Aers,  Phys. Rev. B. {\bf 28}, 1701 (1983).

\bibitem{MRoss79}
Marvin Ross, Phys. Rev B {\bf 21}, 3140 (1980).

\bibitem{Vinko}
S. M. Vinko, O. CiriCosta, J. S. Wark, Nature Com. {\bf 5}, 3533 (2014).

\bibitem{Theye70}
M. L. Th\`{e}ye, Phys. Rev. B {\bf 2}, 3060 (1970).

\bibitem{Ping-Ng06}
Y. Ping, D. Hanson, I. Koslow, T. Ogitsu, D. Prendergast, E. Schwegler,
 G. Collins, and A. Ng, Phys. Rev. Lett. {\bf 96}, 255003 (2006).


\bibitem{RosAsh}
Y. Rosenfeld and N. W. Ashcroft, Phys. Rev A {\bf 20}, 1208 (1979).

\bibitem{Norman68}
G. E. Norman and A. N. Starostin, High Temp. {\bf 6}, 394 (1968).

\bibitem{Nellis96}
S. T. Weir, A. C. Mitchell, and W. J. Nellis, Phys. Rev. Lett. {\bf 76}, 1860 (1996).
\bibitem{Nellis13}
W. J. Nellis, High Pressure Res. {\bf 33}, 369 (2013).

\bibitem{LorenzenRedmer10}
W. Lorenzen, B. Holst, and R. Redmer, Phys. Rev. B {\bf 82}, 195107 (2010).

\bibitem{MoralesCep13}
M. A. Morales, J. M. McMahon, C. Pierleoni, and D. M. Ceperley, Phys. Rev. Lett. {\bf 110}, 065702 (2013).

\bibitem{kedge}
F. Perrot and M. W. C. Dharma-wardana, Phys. Rev. Lett. {\bf 71}, 797 (1993).

\bibitem{Demura2009}
A. V. Demura,  Int. J. of Spec., {\bf 2010}  doi:10.1155/2010/671073 (2009).

\bibitem{IchiIy}
 H. Iyetomi and S. Ichimaru, Phys. Rev. A {\bf 34}, 433 (1986).

\bibitem{prl2}
Francois Perrot and M. W. C. Dharma-wardana, Phys. Rev. Lett. {\bf 87}, 206404 (2001).

\bibitem{SPanDufty}
James Dufty and Sandipan Dutta, Phys. Rev. E {\bf 87}, 032101 (2013).

\bibitem{bulutay2002}
C. Bulutay and B. Tanatar, Phys. Rev. B {\bf 65}, 195116 (2002).

\bibitem{totsuji041
}N. Q. Khanh and H. Totsuji, Solid State Com., {\bf 129}, 37 (2004).

\bibitem{Sjostrom1}
Travis Sjostrom, Frank E. Harris, and S. B. Trickey,
Phys. Rev. B {\bf 85}, 045125 (2012).

\bibitem{eosb}
F. Perrot, M.W.C. Dharma-wardana, and J. Benage, Phys. Rev. E {\bf 65}, 046414  (2002).

\bibitem{HessEbl86}
H. Hess and W. Ebeling,  in 
 Strongly Coupled Plasma Physics, p 185  (edis.) by Forrest J. Rogers and Hugh E. Dewitt (Plenum, New York 1987).

\bibitem{Kraeft86}
D. W. Kr\"{a}ft, D. Kremp, W. Ebeling, and G. R\"{o}pke, Quantum Statistics of Charge Particle Systems,  Plenum Press, New York (1986).

\bibitem{Deutsch73}
C. Deutsch, M. Lavaud, Phys. Lett A {\bf 43}, 193 (1973).

\bibitem{Lado67}
F. Lado, J. Chem. Phys. {\bf 47}, 5369 (1967).


\bibitem{Chihara86}
J Chihara  J. Phys. C: Solid State Phys. {\bf 19}, 1665 (1986).


\bibitem{Kalman97}
G. Kalman, in Strongly Coupled Coulomb Systems, (Eds.),
Gabor J. Kalman, J. Martin Rommel, Krastan Blagoev,
  (Springer, New York 1997)
\bibitem{Ashcroft2002}
Neil Ashcroft, J. Phys. A (SCCS2002), {\bf 36}, 6137 (2002).
\bibitem{HugExp1}
L. B. Da Silva, P. Celliers, G. W. Collins, K. S. Budil, N. C. Holmes,T. W. Barbree, B. A. Hammel, J. D. Kilkenny, R. J. Wallace, M. Ross, R. Cauble, A. Ng and G. Chiu, Phys. Rev. Lett. {\bf 78}, 483 (1997)  pp 331-369

\bibitem{Mahantxt} G. D. Mahan,  Many particle Physics, (Plenum
 New York 1981); J. Phys. Chem. Solids {\bf 31}, 1477
(1970).

\bibitem{PDW-Thermophys}
  Fran\c{c}ois Perrot and M. W. C. Dharma-wardana,
  International Journal of Thermophysics, {\bf 20}, 1299 (1999).

\bibitem{benageR}
J. F. Benage, W. R. Shanahan and M. S. Murillo, Phys. Rev. Lett.
{\bf 83}, 2953 (1999).
\bibitem{desk}
A. W. DeSilva and J. D. Katsouros, Phys. Rev. E. {\bf 57}, 5945 (1998).

\bibitem{kuhl}
S. Kuhlbrodt and R. Redmer, Phys. Rev. E, {\bf 62}, 7191 (2000).

\bibitem{Mazevet07}
S. Mazevet, M. P. Desjarlais, L. A. Collins, J. D. Kress and N. H. Magee,
Phys. Rev. E {\bf 71}, 016409 (2005).

\bibitem{leemore}
 Y. T. Lee and R. M. More, Phys. Fluids, {\bf 27}, 1273 (1984).

\bibitem{dwp-hopping92}
M.W.C. Dharma-wardana and F. Perrot, Phys. Rev. A {\bf 45},5883 (1992).

\bibitem{rein-rop14}
H. Reinholz, G. R\"{o}pke, S. Rosmej, R. Redmer,
http://arxiv.org/pdf/1401.0805.pdf

\bibitem{R-Milsch-Alu92}
H. M. Milchberg, R. R. Freeman, S. C. Davy, and R. M. More,
Phys. Rev. Lett. {\bf 61}, 2364 (1988).
 
\bibitem{DynRes92}
M.W.C. Dharma-wardana and F. Perrot,  Phys. Letters A  {\bf 163},223 (1992).

\bibitem{Furitani}
F. Perrot, Y. Furutani, and M. W. C. Dharma-wardana,
Phys. Rev. A {\bf 41}, 1096 (1990).

\bibitem{HM-md} 
 J.-P Hansen and I. R. McDonald, Phys. Lett. A  {\bf 97}, 42-44 (1983).

\bibitem{Belfast}
D. Riley, N.C. Woolsey, D. McSherry, I. Weaver, A. Djaoui, and E. Nardi, 
Phys. Rev. Lett. {\bf 84}, 1704 (2000).

\bibitem{BJPS}
L. S. Brown, D. L. Preston and R. L. Singleton, Jr.,
 {\it Phys. Rep.} {\bf 410}, 237 (2005).

\bibitem{mmll}
 J. N. Glosli, F. R. Graziani, R. M. More, M. S. Murillo,
F. H. Streitz, M. P. Surh, L. X. Benedict, S. Hau-Riege,
A. B. Langdon, and R. A. London, arXiv:0802.4037.
 Phys. Rev. E {\bf 78}, 025401(R) (2008).


\bibitem{SMVG2010}
M. Schlanges, T. Bornath, J. Vorberger, and Dirk O. Gericke, 
Contributions to Plasma Physics, {\bf 50}, 64-68. (2010).

\bibitem{DD08}
G. Dimonte and J. Daligault, Phys. Rev. Lett. {\bf 101}, 135001 (2008).

\bibitem{Bene12}
Lorin X. Benedict, Michael P. Surh, John I. Castor, Saad A. Khairallah, Heather D. Whitley, David F. Richards, James N. Glosli, Michael S. Murillo, Christian R. Scullard, Paul E. Grabowski, David Michta, and Frank R. Graziani. Phys. Rev. E {\bf 86}, 046406 (2012).

\bibitem{ElrPRL91}
M.W.C. Dharma-wardana,  Phys. Rev. Lett. {\bf 66}, 197 (1991).

\bibitem{ELR-ssc}
M.W.C. Dharma-wardana,  Solid State Com.,
{\bf 86}, 83 (1993).

\bibitem{GiuVig05}
G. Guiliani and G. Vignale,  Quantum Theory of the Electron Liquid, Cambridge, UK (2005).

\bibitem{cdwELR01}
M. W. C. Dharma-wardana,  Phys. Rev. E {\bf 64}, 035432(R) (2001).

\bibitem{PerrotKE}
F. Perrot, Journal of Physics-Condensed Matter
{\bf 6}, 431 (1994). 

\bibitem{TrickeyKE}
Valentin V. Karasiev, Travis Sjostrom, and S. B. Trickey, Phys. Rev. B {\bf 86}, 115101 (2012).

\bibitem{Sash2003}
A. A. Shashkin, M. Rashmi, S. Anissimova, and S. V. Kravchenko, V. T. Dolgopolov,
T. M. Klapwijk, Phys. Rev. Lett. {\bf 91}, 46403 (2003).

\bibitem{2v2d}
M. W. C. Dharma-0wardana and Fran\c{c}ois Perrot, Phys. Rev. B {\bf 70}, 035308 (2004)

\bibitem{dePaloSan}
M. Marchi, S. De Palo, S. Moroni, and Gaetano Senatore. Phys. Rev. B {\bf 80}, 035103 (2009).

\bibitem{Laughlin}
R. B. Laughlin, Phys. Rev. Lett. {\bf 50}, 1395 (1983).

 \bibitem{JancoviciFQHE}
B. Jancovici, Phys. Rev. Lett. {\bf 46}, 386 (1981).

\bibitem{CDW-Aers}
M. W. C. Dharma-wardana and G. C. Aers, unpublished.

\bibitem{Gorbig11}
M. O. Goerbig, Rev. Mod. Phys. {\bf 83}, 1193  (2011).

\bibitem{Nielson-Perali-13}
A. Perali, D. Nielson, A. R. Hamilton, 
Phys. Rev. Lett. {\bf 110}, 146803 (2013).

\end{thebibliography}
\end{document}